**Revisiting XDoppler estimator for high spatiotemporal resolution volumetric axial velocity measurement using row-column arrays**

Henri Leroy[1], Adrien Bertolo[1], Guillaume Goudot[2], Mickaël Tanter[1], Thomas Deffieux[1], Mathieu Pernot[1,*]

[1] Physics for Medicine Paris, INSERM U1273, ESPCI Paris, PSL University, CNRS, Paris, France

[2] Vascular medicine department, Hôpital Européen Georges Pompidou, Assistance Publique-Hôpitaux de Paris (AP-HP), Paris, France

[*]Corresponding author: Mathieu Pernot (mathieu.pernot@espci.fr)

**Abstract**

Accurate volumetric velocity estimation is crucial in ultrasound imaging for both diagnostic and therapeutic applications. Traditional ultrasound systems, though effective for two-dimensional imaging, face major limitations in 3D imaging due to hardware and computational demands. Row-column addressed (RCA) ultrasound probes offer a promising alternative by reducing hardware complexity, thereby reducing the gap between research prototypes and clinical systems. However, this typically comes at the expense of stronger sidelobes compared with fully populated matrix arrays, leading to reduced image contrast. Several approaches have been proposed to improve the contrast of power Doppler imaging, yet the accuracy and performance of velocity Doppler estimation have received comparatively little attention. In this study, we present a method that exploits the phase information from RCA row and column signals to derive a novel velocity estimator based on cross-correlation of orthogonal apertures. This extends the XDoppler scheme, initially developed for power Doppler imaging, to velocity estimation. The XDoppler estimator is shown to provide accurate measurements of axial velocities and to outperform the traditional phase-shift autocorrelator, while offering a theoretical Nyquist velocity twice as high. *In vitro* experiments further demonstrate enhanced sensitivity to slow flows and reduced bias in flow rate estimation. *In vivo* data from a carotid artery confirm the reduced sensitivity to aliasing and reveal the ability to track dynamic blood flow velocity changes associated with arterial pulsatility. These findings suggest that the XDoppler velocity estimator could improve volumetric velocity imaging in clinical contexts.

## 1. Introduction

Ultrafast ultrasound imaging has become an invaluable tool for a wide range of applications, including tissue elasticity imaging [1], [2], functional imaging [3], blood flow dynamics at macroscopic and microscopic scales [4], [5], [6], [7], [8], [9], [10]. Despite these advances, achieving accurate three-dimensional (3D) ultrafast imaging with ultrasound systems remains challenging due to the substantial hardware and computational demands. Traditional, fully populated matrix arrays for 3D imaging often require extensive channel interconnections, leading to increased system size, cost, and power requirements, which creates barriers to widespread use in portable or wearable applications.

In recent years, Row-Column Addressed (RCA) ultrasound probes have emerged as a promising alternative to fully-populated matrix arrays for ultrafast imaging. By addressing rows and columns separately, RCA probes significantly reduce hardware complexity and the number of required channels and associated electronics [11], [12], [13], [14], [15]. Since the original design, various configurations have been proposed, such as curved probes [16] or flat probes with a lens [17] enabling these arrays to emit diverging waves.

With conventional hardware, RCA arrays can capture large volumes while maintaining a reasonably high framerate [18], [19]. As a result, RCA arrays have been investigated for a variety of applications, such as Doppler imaging [20], [21], [22], [23], [24], [25], [26], [27], [28], ultrasound localization microscopy (ULM) [29], [30], [31], [32] or shear wave elastography [33], [34]. All of these techniques require an accurate estimation of motion to provide accurate and quantitative information that can be used in clinical practice. Yet, the RCA probe architecture also introduces specific challenges compared to matrix arrays, such as strong side lobes leading to a poorer spatial resolution and a poorer contrast [15]. In many situations, this can result in an inaccurate motion estimation [20], [35]. Therefore, robust and accurate estimations of motion using RCA probes remain an area of investigation.

To tackle these challenges, different imaging strategies have been proposed, such as synthetic aperture (SA) imaging [12], [13] or plane wave (PW) imaging [15], [20]. In plane wave imaging, Orthogonal plane wave (OPW) compounding was first introduced, based on coherent compounding of two sub-volumes acquired by successively emitting with rows and receiving with columns and vice-versa, to reconstruct an isotropic point spread function (PSF) [1], [2], [15], [20]. In recent years, alternatives to OPW have been proposed to improve image quality, contrast, and resolution. The XDoppler method, which is based on cross-correlation as opposed to coherent compounding of orthogonal apertures, has been proven to decrease sidelobes and enhance signal-to-noise ratio for power Doppler imaging [21]. Other approaches have been proposed based on the same concept, such as the row-column frame multiply and sum (RC-FMAS) scheme [36], the X frame multiply and sum (X-FMAS) [37] or spatio-temporal similarity weighting (St-SW) [38]. Most of these works have focused on improving contrast and resolution; however, their potential to enhance the estimation of tissue motion and blood flow has received little attention.

In the present work, we revisit the concept of cross-correlation of orthogonal apertures introduced with XDoppler [21] for power Doppler imaging to derive a novel high-spatiotemporal-resolution estimator of axial velocity for volumetric imaging using row-column addressed arrays. We first demonstrate the need for a new velocity estimator through ultrasound simulations, then provide a mathematical derivation of the proposed Doppler velocity estimator. Experimental validation is subsequently performed *in vitro* using a flow phantom and *in vivo* on the carotid artery of a healthy volunteer, thus introducing a new color Doppler solution for RCA arrays.

## 2. Methods

### 2.1 Theoretical considerations

#### 2.1.1 Image formation with a Row-Column Addressed array

To achieve a volumetric image with a RCA array, the two main acquisition methods used are synthetic aperture imaging (SAI) [12], [13] and plane wave imaging (PWI) [15], [20], which both consist of acquiring consecutively multiple low-resolution volumes with different transmits (thus sampling the volume with various wave vectors, which then allows to perform synthetic focusing in emission) and combining them to create one final high-resolution volume. For plane wave imaging, the usual method consists in transmitting multiple plane waves with the rows (R) and receiving with the columns (C). After I/Q demodulation, beamforming and coherent compounding of the plane waves, this leads to the formation of a first sub-volume called « RC » and represented by the complex signals $s_{RC}$. Then, multiple plane waves are transmitted with the columns (C) and received with the rows (R). After I/Q demodulation, beamforming and coherent compounding of the plane waves, this leads to the formation of another sub-volume called « CR », represented by the complex signals $s_{CR}$. An example of an ultrafast acquisition sequence with alternate transmissions by R and C arrays to acquire RC and CR sub-volumes is given in Figure 1.

Figure 2 displays the PSFs obtained by plane wave imaging with a RCA array through a Field II [39] simulation of a static single scatterer, imaged with 10 + 10 plane waves (PRF = 12 kHz, transmit frequency 5 MHz) using a 128 + 128-element RCA array (center frequency 6 MHz, aperture area 25.6×25.6 mm²): because of synthetic focusing in transmission, the PSF of RC (2.a) and CR (2.b) are elongated in the direction of the array transmitting the plane waves: the PSF of RC is therefore the same as the PSF of CR, with a rotation of $\frac{\pi}{2}$.

To retrieve a more isotropic PSF, the two sub-volumes are then compounded to form a volumetric image. This compounding can be a coherent summation in the case of Orthogonal Plane Wave compounding (OPW scheme) [15], a compounding based on cross-correlation (XDoppler scheme) [21], frame-multiply-and-sum compounding (RC-FMAS or X-FMAS) [36], [37], or spatio-temporal similarity weighting scheme (St-SW) [38].

Neglecting the time delay between the acquisition of the two sub-volumes, the OPW compounding operation can be written as:

$$s_{OPW}(i) = s_{RC}(i) + s_{CR}(i) \quad (1)$$

where i represents the index of the samples on the slow-time axis. The PSF of OPW is represented in Figure (2.c) and highlights the impact of the coherent summation operation on the image.

The XDoppler compounding, which was initially introduced for Power Doppler imaging [16] is based on the correlation product of $s_{RC}$ and $s_{CR}^*$:

$$s_X = \frac{1}{N}\sum_{i=1}^{N} s_{RC}(i)s_{CR}^*(i) \quad (2)$$

where i represents the index of the samples on the slow-time axis and N represents the ensemble length of integration of XDoppler imaging. The PSF of XDoppler is represented in Figure (2.d): its full-width at half maximum (FWHM) is smaller than with OPW compounding (1.21 vs. 2.16mm), and the sidelobes level of XDoppler is 4dB lower than OPW thanks to the cross-correlation operation: XDoppler thus provides a better lateral resolution and contrast than OPW, as described in earlier literature [21].

Since the introduction of XDoppler, other methods have tackled this issue of sidelobes, such as the RC-FMAS, X-FMAS, and St-SW schemes [36], [37], [38]. All of these techniques have been developed to decrease the amplitude of the sidelobes to enhance the PSF for ULM applications or power Doppler imaging, but their phase information has not been exploited specifically for velocity estimation, especially because all these methods do not retain the phase information in the same way.

To investigate what information the phase of the signals could provide for velocity estimation, it can be noted that the acquisition of the RC and the CR sub-volumes is not simultaneous. The number of plane waves transmitted by one array (either R or C) is called $N_{angles}$, and $T = 2N_{angles}T_{PRF}$ is the total time needed to form a volumetric image. Then, if $N_{angles}$ plane waves are emitted with the rows successively, and then $N_{angles}$ plane waves are emitted with the columns, as shown in Figure 1, the temporal delay between the formation of a RC sub-volume $s_{RC}(i)$ and the corresponding CR sub-volume $s_{CR}(i)$ is $\Delta T_1 = N_{angles}T_{PRF}$ and the temporal delay between the formation of a CR sub-volume $s_{CR}(i)$ and the next RC sub-volume $s_{RC}(i+1)$ is $\Delta T_2 = N_{angles}T_{PRF} = \Delta T_1$.

Due to this time delay, the CR sub-volume of index i will be referred to as $s_{CR}(i+\Delta T_1/T)=s_{CR}(i+0.5)$ to explicitly indicate this time lag.

Table 1 summarizes the various notations used to describe these quantities throughout the article.

Table 1: Notations and symbols

| | |
|---|---|
| $T_{PRF}$ | Time between two pulse emissions |
| T | Total time to form a volumetric image |
| $N_{angles}$ | Number of plane waves transmitted by one array (R or C) to form a volumetric image |
| $\Delta T_1$ | Time between a RC sub-volume and the corresponding CR sub-volume |
| $\Delta T_2$ | Time between a CR sub-volume and the next RC sub-volume |

### 2.1.2 Conventional Autocorrelation Phase-based axial velocity estimation

In 1985, Kasai et al. introduced a phase-shift axial velocity estimator based on the autocorrelation of the slow-time complex signals [28] :

$$v_z = -\frac{c}{4\pi f_0}\frac{1}{T_{PRF}}\arg\left(\sum_{i=1}^{N-1} s(i)s^*(i+1)\right) (3)$$

where $v_z$ is the axial velocity, c the speed of sound in the medium, $f_0$ the central frequency of the pulse, s(i) the complex signal in a given pixel of coordinates (x,z) for the temporal sample i, $T_{PRF}$ is the time associated with the pulse repetition frequency, N is the ensemble length for Doppler integration, and arg(.) represents the argument of a complex number. The demonstration of this formula can be performed using a simplified model of the slow-time complex signals described in [40], [41], [42], [43]. The slow-time complex signals can be expressed as:

$$s(i) = A_s(i)e^{j(i\Delta\varphi_{Doppler} + \varphi_s)}$$

with the Doppler phase-shift $\Delta\varphi_{Doppler} = \frac{4\pi f_0 v_z}{c}T_{PRF}$ with i the slow time sample. This model is valid under certain approximations detailed in [44]. The Kasai estimator can be computed as the lag-1 autocorrelation as follows:

$$s(i)s^*(i+1) = A_s(i)A_s(i+1)e^{-j\Delta\varphi_{Doppler}}$$

This leads to:

$$\Delta\varphi_{Doppler} = -\arg\big(s(i)s^*(i+1)\big)$$

Thus:

$$v_z = -\frac{c}{4\pi f_0}\frac{1}{T_{PRF}}\arg\big(s(i)s^*(i+1)\big)$$

Averaging over time leads to the Kasai velocity estimator:

$$v_z = -\frac{c}{4\pi f_0}\frac{1}{T_{PRF}}\arg\left(\sum_{i=1}^{N-1} s(i)s^*(i+1)\right)$$

This estimator evaluates the axial velocity, i.e., the projection of the velocity vector in the probe axial direction $\boldsymbol{u_z}$. If there is an angle $\theta$ between the ultrasound beam and the velocity vector $\boldsymbol{v}$ (beam-to-flow angle), one can write: $v_z = \boldsymbol{v}\cdot\boldsymbol{u_z} = v\cos\theta$.

In this context, since the argument is constrained in the interval [-π, π], the Nyquist velocity is given by:

$$v_N = \frac{c}{4f_0 T_{PRF}} \quad (4)$$

The autocorrelator introduced in (3) has been proven to be a relatively accurate estimator of the mean axial velocity of the blood flow. This estimator and some adapted versions [45] have since then widely been used for color Doppler imaging in both research and clinical scanners due to their computational efficiency and ease of implementation.

#### 2.1.3 Axial velocity estimation using orthogonal plane wave imaging

Using plane wave compounding with a RCA array, the phase-shift autocorrelation estimator can be written:

$$v_z = -\frac{c}{4\pi f_0}\frac{1}{T}\arg\left(\sum_{i=1}^{N-1} s_{comp}(i)s_{comp}^*(i+1)\right) \quad (5)$$

with $s_{comp}$ the complex signals in a given voxel of coordinates (x,y,z) corresponding to the compounded volumes (the compounding method being any method that preserves the phase information, such as OPW or RC-FMAS).

Using (4), the theoretical Nyquist velocity for this processing scheme is then:

$$v_N^{comp} = \frac{c}{4f_0 T} = \frac{c}{8f_0 N_{angles} T_{PRF}} \quad (6)$$

A major limitation of this velocity estimator is the low spatial resolution and contrast associated to OPW coherent compounding because of the high side lobes [14]. To illustrate this effect, the image of a single scatterer undergoing displacement was simulated. The OPW-compounded plane wave sequence used in Figure 1 was applied, with the scatterer being moved away from the probe at $v_z$=12mm/s between each transmission/reception event. The resulting phase shift, induced by the scatterer's motion, extends several millimeters beyond its actual position, thereby limiting spatial resolution, contrast, and potentially leading to inaccurate velocity estimates, as shown in Figure 3.

#### 2.1.5 Revisiting the XDoppler estimator for axial velocity estimation

The XDoppler approach was initially proposed to improve the contrast and resolution of power Doppler images by taking advantage of blood signal decorrelation on the orthogonal apertures of RCA probes [21]. Here, we propose to extend the XDoppler approach to axial velocity estimation, broadening the abilities of this method to color Doppler imaging.

XDoppler relies on the cross-correlation between the two sub-volumes RC and CR, as evidenced in (2). By analogy with the Kasai autocorrelation estimator, one could then simply try to extract the phase of the XDoppler cross-correlation to estimate the axial velocity, i.e., consider:

$$\arg\left(\sum_{i=1}^{N} s_{RC}(i)s_{CR}^*(i+0.5)\right) \quad (7)$$

However, unlike the phase of the autocorrelation with the Kasai estimator, $s_{RC}$ and $s_{CR}$ are acquired using orthogonal arrays, which introduces a bias in the phase through the contributions of clutter and side lobes to the phase signal, as shown in Figure 1 and Figure 4. The origin of the bias is demonstrated below.

Using the simplified model introduced to demonstrate the Kasai estimator for both signals acquired using RC and CR schemes, a new XDoppler velocity estimator can be derived. Considering the slow-time complex signals described in 2.1.1 applied to RC and CR in a given voxel of coordinates (x,y,z) [40], [41], [42], [43], the following equations can be written:

$$\begin{cases} s_{RC}(i) = A_{RC}(i)e^{j(i\Delta\varphi_{Doppler} + \varphi_{RC})} \\ s_{CR}(i+0.5) = A_{CR}(i+0.5)e^{j((i+0.5)\Delta\varphi_{Doppler} + \varphi_{CR})} \end{cases}$$

The Doppler phase shift $\Delta\varphi_{Doppler} = \frac{4\pi f_0 v_z}{c}T$ is assumed to be identical for both signals (because RC and CR are images of the same voxel). Importantly, the phases $\varphi_{RC}$ and $\varphi_{CR}$ are, however, specific to each array and depend on the chosen voxel (x, y, z). Assuming there is no decorrelation (which will be discussed later), $\varphi_{RC}$ and $\varphi_{CR}$ are supposed to be constant between $s_{RC}(i)$ and $s_{RC}(i+1)$.

Then, two cross-correlations $C_1=s_{RC}(i)s_{CR}^*(i+0.5)$ and $C_2=s_{CR}(i+0.5)s_{RC}^*(i+1)$ can be computed:

$$\begin{cases} C_1 = A_{RC}(i)A_{CR}(i+0.5)e^{j(-0.5\Delta\varphi_{Doppler} + \varphi_{RC}-\varphi_{CR})} \\ C_2 = A_{CR}(i+0.5)A_{RC}(i+1)e^{j(-0.5\Delta\varphi_{Doppler} + \varphi_{CR}-\varphi_{RC})} \end{cases}$$

This leads to:

$$\begin{cases} \arg(C_1) = -0.5\Delta\varphi_{Doppler} + \varphi_{RC} - \varphi_{CR} \\ \arg(C_2) = -0.5\Delta\varphi_{Doppler} + \varphi_{CR} - \varphi_{RC} \end{cases}$$

This system of equations explains why computing only one cross-correlation leads to a biased estimator, as the Doppler phase-shift cannot be isolated, which leads to the complex spatial patterns shown in Figure 1 and Figure 4.

The two phases can then be summed together to disentangle them:

$$\arg(C_1) + \arg(C_2) = -\Delta\varphi_{Doppler}$$

Averaging over time is finally performed, just like for the Kasai estimator, and provides the combined XDoppler axial velocity estimator:

$$v_z = -\frac{c}{4\pi f_0}\frac{1}{T}\left[\arg\left(\sum_{i=1}^{N-1} s_{RC}(i)s_{CR}^*(i+0.5)\right) + \arg\left(\sum_{i=1}^{N-1} s_{CR}(i+0.5)s_{RC}^*(i+1)\right)\right] \quad (8)$$

Figure 4.c shows how it improves the spatial resolution and contrast of the phase estimation for one moving scatterer compared to the one obtained by OPW (Figure 3).

This new estimator also extends the Nyquist velocity by subdividing the total observation time $T(\Delta T_1=\Delta T_2=T/2)$. While the conventional Kasai estimator relies on a single autocorrelation over T, this method utilizes two separate cross-correlations over shorter intervals ($\Delta T_1=\Delta T_2=T/2$). Because the unambiguous velocity range is constrained by the 2π periodicity of the phase shift, halving the effective displacement time effectively doubles the Nyquist limit:

$$v_N^X = \frac{c}{2f_0T} = \frac{c}{4f_0N_{angles}T_{PRF}} = 2v_N^{comp} \quad (9)$$

### 2.2 Acquisitions and processing pipeline

Ultrasound imaging was performed using a Verasonics Vantage 256 High Frequency ultrasound scanner equipped with a Vermon 128+128 6MHz transducer array (RC6gV): the characteristics of the probe are detailed in Table 2.

Table 2: Probe specifications

| | |
|---|---|
| Number of elements | 128+128 |
| Central frequency (MHz) | 6 |
| Pitch (mm) | 0.2 |
| Kerf (mm) | 0.025 |
| Bandwidth (%) | 100 |
| Aperture (mm$^2$) | 25.6×25.6 |

The characteristics of the imaging sequences used are described in Table 3: 2*$N_{angles}$ tilted plane waves ($N_{angles}$ per subarray) of central frequency $f_c$ were sent at a given pulse repetition frequency, corresponding to an effective volume-rate of $\frac{PRF}{2N_{angles}}$. The imaging pulse duration was 2 cycles with a pulse duty cycle of 1. The angular step was 0.5° to reject angular grating lobes out of the field of view, according to the theory on angular sampling developed for synthetic focusing using plane wave imaging [1], [20], [46]. The speed of sound is assumed to be constant (c=1540m/s).

Table 3: Imaging sequences

| | Flow phantom – Experiment 1 (Φ = 8mm) | Flow phantom – Experiment 2 (Φ = 4mm) | Human carotid |
|---|---|---|---|
| PRF (Hz) | 12000 | 20000 | 12000 |
| Number of plane waves | 10+10 | 8+8 | 10+10 |
| Effective volume rate (Hz) | 600 | 1250 | 600 |
| Central frequency (MHz) | 5 | 6.25 | 5 |
| Nyquist velocity - Autocorrelation (cm/s) | 4.6 | 7.7 | 4.6 |
| Nyquist velocity - XDoppler (cm/s) | 9.2 | 15.4 | 9.2 |
| Angular range (°) | 4.5 | 3.5 | 4.5 |
| Angular step (°) | 0.5 | 0.5 | 0.5 |
| Ensemble length (ms) | 500 | 112 | 4000 |

The signals were sampled on all channels at 4$f_c$ sampling rate and 100% bandwidth. Data were processed offline using MATLAB (2023a, MathWorks, Natick, MA, USA). After in-phase and quadrature demodulation, the signals corresponding to each plane wave were beamformed using a delay and sum algorithm implemented in CUDA language and run on a GPU (Nvidia GeForce RTX 3070 Ti) to form a low-resolution volume, with a voxel size of 0.2mm×0.2mm×0.2 mm. The $N_{angles}$ volumes corresponding to RC were coherently compounded to form the RC signals $s_{RC}$ and the $N_{angles}$ volumes corresponding to CR were coherently compounded to form the CR signals $s_{CR}$.

The compounded RC and CR signals were filtered separately using spatiotemporal singular value decomposition to remove the tissue signal [47]. In the *in vitro* experiments, 5% of the singular values were removed for filtering to provide effective static tissue and clutter suppression while preserving the flow signal. In the *in vivo case,* clutter filtering was performed using an adaptive method based on the spatial similarity matrix [48].

XDoppler compounding and OPW compounding were then performed separately. The correlations for velocity estimation were computed and averaged over a spatial window of 3×3×3 voxels before temporal integration. Phase computation and axial velocity estimation were then performed based on (5) and (8). Power Doppler estimations were also computed for OPW and XDoppler schemes (the XDoppler power Doppler is computed after a square root operation to account for the correlation product and be expressed in the same units as the OPW power Doppler).

The velocity maps are masked based on the XDoppler Power Doppler with a 15dB threshold for visualization purposes.
The whole processing pipeline is summarized in Figure 5.

## 2.3 Experimental setups

### 2.3.1 *In vitro* experiments

A flow phantom (Model 523A, ATS Laboratories, Bridgeport, CT, USA) was used to simulate blood flow within a vessel-like structure with an internal diameter of 8mm or 4mm, depending on the experiments.
The channel was connected to a closed-loop flow system comprising a reservoir and a peristaltic pump with a controllable flow rate (MasterFlex L/S, Model 07528-10, Cole-Parmer Instrument Company, Antylia Scientific, IL, USA), and tubing to create a steady-state circulation with a blood-mimicking fluid made out of degassed water and cellulose particles [49], [50], [51]. The probe was positioned on the top of the phantom. The tube vessel was centered in the field of view using a real-time biplane B-mode imaging solution implemented directly on the RCA probe, displaying orthogonal YZ and XZ slices. The angle between the normal to the probe surface and the tube axis (i.e., the beam-to-flow angle) is $\theta = 72°$. Various flow rates were used to assess the performance of the estimators in different flow conditions, during two separate experiments, as shown in Table 3.
In experiment 1, 6 different flow rates were used to first assess the accuracy of the XDoppler estimator and then compare it with the autocorrelation estimator while being under the aliasing limit ($Q_1 = 125 mL/min$, $Q_2 = 74 mL/min$, $Q_3 = 57 mL/min$, $Q_4 = 71 mL/min$, $Q_5 = 93 mL/min$, and $Q_6 = 114 mL/min$ ). Experiment 2 was designed to evaluate the robustness of the XDoppler estimator to aliasing, and demonstrate experimentally that its Nyquist limit is higher than the autocorrelation estimator: 2 different flow rates were used to evaluate the robustness of the estimator below and above the aliasing limit of the Kasai estimator (which was $v_N^{comp} = 7.7\ cm/s$ ) while staying below the aliasing limit of the XDoppler estimator (which was $v_N^X = 15.4\ cm/s$). The chosen flow rates were $Q_a = 37 mL/min$, corresponding to a maximal axial velocity of $v_{z,a} = 3\ cm/s$ and $Q_b = 146 mL/min$, corresponding to a maximal axial velocity of $v_{z,b} = 12\ cm/s$. These flow rates can correspond to flow rates that are found in carotid arteries of healthy or stenosed patients [52], [53], [54].

### 2.3.2 *In vivo* human carotid

Acoustic parameters of the sequence were measured using a calibrated interferometer in degassed water [55]. The measured central pulse frequency was 4.88MHz, deviating by 2.4% from the nominal central pulse frequency. Because of this small discrepancy, the nominal frequency was used for the processing.
For a voltage U=60V at a PRF of 12kHz, the mechanical index (MI) was 0.43, the thermal index in soft tissue (TIS) was 0.76, the spatial-peak temporal average (ISPTA) was 162 mW/cm$^2$ and the spatial-peak pulse average (ISPPA) was 25 W/cm$^2$, in compliance with the recommendations of the Food and Drug Administration (FDA 510k) [56]. Surface heating of the probe was measured over 30 minutes with a sequence designed to yield an effective duty cycle of 1/3, alternating between 4-s acquisitions and 8-s pauses, and complied with the IEC regulation (IEC 60601-2-37) [56].

Acquisitions were performed on the carotid artery of a healthy volunteer after obtaining his informed consent in the context of a clinical study approved by the French ethical committee (ClinicalTrials.gov Identifier: NCT04470687). Using the real-time B-mode imaging mode implemented on the RCA probe, the probe was positioned above the carotid artery and jugular vein of the volunteer, after which a 4s acquisition corresponding to the calibrated sequence was initiated.

### 2.4 Evaluation of the estimator and associated metrics

To evaluate the performance of the XDoppler estimator on *in vitro* data, a Poiseuille model for a cylindrical pipe of radius R and a laminar flow (Reynolds number : $Re \sim \frac{\rho_{water} v_{max} R}{\eta} \sim 10^2$) was selected, with a velocity profile given by $\vec{v}(r) = v_{max}\left(1 - \left(\frac{r}{R}\right)^2\right)\boldsymbol{u_{axis}}$, where r is the radial coordinate with respect to the pipe axis, $\boldsymbol{u_{axis}}$ the axis of the pipe cylinder and $v_{max}$ the velocity at the center of the pipe. Under this model, the flow rate dispensed by the pump is linked to the velocity field by $Q_{dispensed} = \iint_S \boldsymbol{v} \cdot \boldsymbol{dS} = \pi R^2 v_{mean} = \pi R^2 \frac{v_{max}}{2}$ with $v_{mean}$ the average flow velocity. Evaluating the flow rate derived from the velocity is a way of averaging the values over all the voxels, thus making it more robust to noise and closer to an accurate description of the global flow dynamics.

Several metrics are used to evaluate the quality of the velocity estimation of each method:

1) Root mean squared error: $RMSE = \frac{1}{v_{Poiseuille}^{max}} \sqrt{\frac{1}{N_{pixel}} \sum_{i=1}^{N_{pixel}} (v_{est}(x,y,z) - v_{Poiseuille}(x,y,z))^2}$ where $N_{pixel}$ is the number of imaging pixels in the vessel, $v_{est}$, and $v_{Poiseuille}$ are the estimated velocity and theoretical velocity according to the Poiseuille model, respectively, and $v_{Poiseuille}^{max}$ is the peak theoretical velocity of the simulated vessel.
2) Estimated volumetric flow rate: the volumetric flow rate through a given surface was estimated based on the flow of the estimated velocity field through a transversal slice of the pipe with a 15dB mask on Power Doppler data.
3) Time of convergence: The time of convergence is related to the ensemble length in Doppler imaging [28], [44], i.e. a time of integration necessary for the estimator to stabilize around its expected value. For each curve, the time of convergence of the estimator $t_{90\%}$ was defined as the duration of acquisition that must be accumulated until 90% of the final flow rate value is reached.
4) Relative error on the flow rate: $e = \frac{|Q_{estimated} - Q_{dispensed}|}{Q_{dispensed}}$
5) Bias on the flow rate: $\bar{B} = \sum_{i=1}^{N} (Q_{dispensed}(i) - Q_{estimated}(i))$ where N is the total number of flow rates investigated (N=6 in experiment 1 and N=2 in experiment 2) and i is the index of a given flow rate
6) Standard deviation on the flow rate: $\bar{\sigma} = \sqrt{\frac{1}{N} \sum_{i=1}^{N} (Q_{estimated}(i) - Q_{dispensed}(i))^2}$ where N is the total number of flow rates investigated (N=6 in experiment 1 and N=2 in experiment 2) and i is the index of a given flow rate

To display the results, several representations were used to analyze the velocity fields and the flow rates:

- volumetric images and cross-sections of the axial velocity field for visual evaluation
- estimation of the radial velocity profile: the central point of the vessel was manually selected, and the radial profile was averaged on multiple radii (N=64) in evenly spaced directions to get a robust estimation of the average velocity profile.
- Voxel-by-voxel comparison for the velocity field: the data were binned into a 200×200 grid, and the number of points in each bin was computed and displayed using a logarithmic scale. A 1:1 reference line was included to facilitate visual comparison between the predicted values and the actual values. Linear regression was also performed to evaluate the agreement between the estimated values and the theoretical model.
- Comparison of the estimated and the theoretical flow rates (including a linear regression and a 1:1 reference line)
- Bland-Altman plot for the flow rates to compare estimated and theoretical values

## 3. Results

### 3.1 *In* vitro flow phantom

#### 3.1.1 Reconstruction of an accurate velocity profile & Comparison with state-of-the-art axial velocity estimator

Figure 6 represents the estimated axial velocities inside a pipe for one dataset of experiment 1. The flow rate dispensed by the pump was $Q_1 = 125mL/min$ and the diameter of the pipe was 8mm. Integration time was 500ms.

Figure 6(a) demonstrates that the axial velocity can be estimated over the entire field of view. Figure 6(b) shows a transverse cross-section of the pipe, illustrating the spatial variations of the velocity field.

Figure 6(c) shows that XDoppler velocity profile follows closely the Poiseuille parabolic profile corresponding to the dispensed flow rate. The velocity is accurately estimated in the central region of the pipe, while the standard deviation of the XDoppler estimator increases near the pipe wall, reflecting reduced sensitivity to very slow flows.

Compared with a conventional axial velocity estimator - the Kasai “lag-1” autocorrelator [28], which was applied to the coherently compounded signals (i.e., the OPW scheme [1], [15]), XDoppler velocity profile visibly follows more closely the theoretical profile. In particular, XDoppler estimates correctly the velocity at the centre of the pipe and around the boundaries, while the Kasai estimator applied to OPW underestimates the central velocity by approximately 10% and overestimates the distal velocity, as the mean velocity does not reach values below 10mm/s whereas the Poiseuille model drops to zero at the boundary of the pipe.

Figure 6(d) compares, for all voxels of the volume, the estimated values against those predicted by a Poiseuille flow profile, for both the XDoppler (d.1) and Kasai autocorrelation (d.2) estimators. The agreement metrics listed in Table 4 are derived from these distributions:

- The XDoppler distribution follows more closely the 1:1 reference line than the autocorrelation estimator, indicating a more accurate estimation
- According to the distributions of the estimates, Kasai autocorrelation is more limited than the XDoppler estimator in the range of velocities that it can estimate, both regarding higher velocities (no estimate above 26mm/s for the autocorrelation, whereas the maximal value estimated by XDoppler can go up to 30mm/s) and lower velocities (no estimate below 7mm/s whereas the minimal value estimated by XDoppler goes down to 4mm/s)

Table 4 displays the agreement between the estimated values and the values predicted by the Poiseuille model with one low flow rate ($Q_2$=74mL/min) and one high flow rate ($Q_1$=125mL/min).

Table 4: Comparison of agreement metrics with Poiseuille model for both estimators for two different flow rates

| | Kasai autocorrelation estimator | XDoppler estimator | |
|---|---|---|---|
| Root mean squared error - RMSE (%)<br>Coefficient of determination ($R^2$) | 11.41<br>0.62 | 9.21<br>0.88 | **$Q_1$=125mL/min** |
| Root mean squared error - RMSE (%)<br>Coefficient of determination ($R^2$) | 12.51<br>0.58 | 8.22<br>0.84 | **$Q_2$=74mL/min** |

The agreement is better for XDoppler than for Kasai estimator for both flow rates:

- RMSE is lower for XDoppler than for Kasai estimator
- The coefficient of determination of the linear regression between the predicted values and the estimated values is higher for XDoppler than for Kasai estimator

Finally, Figure 7(a) and 7(b) display the flow rate estimations for all the experimental conditions evaluated in experiment 1 and shows the agreement between dispensed and estimated flow rates after 0.5s of integration for 6 different flow rates. The Kasai estimator applied to OPW globally overestimates the flow rate with a regression slope of 1.20, a mean bias $\overline{B_{Kasai}} = -9.33$ mL/min and a standard deviation $\overline{\sigma_{Kasai}} = 6.19$mL/min, while XDoppler is closer to the dispensed flow rate with a regression slope of 1.08, a mean bias $\overline{B_{XDoppler}} = -0.17$ mL/min and a standard deviation $\overline{\sigma_{XDoppler}} = 3.43$mL/min.

#### 3.1.2 Flow rate estimation based on axial velocity estimation: accuracy and convergence time

As detailed in previous literature [21], the XDoppler cross-correlation method involves an integration time before stabilization, which can be interpreted as the Doppler ensemble length of the imaging technique [28], [44]. The time of convergence (i.e. the minimal time of integration needed for the estimation to be stable) of the estimator was evaluated in two different flow conditions of experiment 1 (dispensed flow rates of $Q_1 = 125mL/min$ and $Q_2 = 74mL/min$): XDoppler estimator takes between 0.05s and 0.2s to reach $t_{90\%}$. The final error is inferior to 10% (respectively +0.4% and -7.6% for $Q_1$ and $Q_2$), as exposed in Figure 8(a).

Figure 8(b) shows that, for all the flow rates (N=6) of experiment 1, ranging from 57 mL/min to 125 mL/min, the mean relative error over all the flow rates reaches the 10% threshold after around 0.17s of integration and drops down to 2.9% after 0.5s of integration.

#### 3.1.3 Sensitivity to aliasing effects

The robustness to aliasing of the two axial velocity estimators based on (5), (6), (8), and (9) was assessed and compared in the experimental conditions corresponding to experiment 2. In this case, the Nyquist velocity is equal to 7.7 cm/s with the autocorrelation estimator, when it is equal to 15.4 cm/s with the XDoppler estimator.

A first estimation was made at a low flow rate corresponding to an axial velocity of 3 cm/s (Figure 9.a). In this case, the resulting maps show velocity values with the same order of magnitude between the two approaches, and no aliasing is observed.

Then, a second estimation was performed at a higher flow rate, corresponding to an axial velocity of 12 cm/s (Figure 9.b). In this case, aliasing is observed with the autocorrelation estimator, while XDoppler still yields the correct estimation.

### 3.2 *In vivo* human carotid

Acquisition was performed on the common carotid artery of a healthy volunteer. In Figure 10, the 3D reconstructions for the OPW and XDoppler approaches are presented during systole and diastole. One can recognize for both reconstructions the carotid artery and the jugular vein in the field of view.

In Figure 10.a, during diastole, aliasing is observed inside the jugular vein with the autocorrelation estimator because of the geometry of the vein (which goes upwards), while it is not the case with the XDoppler estimator. In Figure 10.b, during systole, aliasing is observed inside the carotid with both autocorrelation and XDoppler estimators during because of the high increase in velocity during systolic ejection. These results confirm that the XDoppler estimator is less sensitive to aliasing effects than the autocorrelation estimator. Moreover, the average velocity measurements inside the vessels were the following:

- the mean value of the axial velocity inside the carotid was almost identical with both methods (0.88cm/s with XDoppler vs. 0.89cm/s with OPW) in diastole while some difference appeared during systole (1.54cm/s with XDoppler vs. 1.32cm/s with OPW), probably due to stronger aliasing in OPW case, leading to a decrease of the absolute value of the velocity.
- the mean value of the axial velocity inside the jugular vein was -0.91cm/s with XDoppler vs. -0.60cm/s with OPW in diastole and -0.46cm/s with XDoppler vs. -0.82cm/s with OPW during systole. The decrease in absolute value of the mean velocity in the jugular during systole with both methods is probably also due to aliasing.

Figure 11 displays the temporal evolution of the average axial velocity inside the carotid artery during 4s using the XDoppler estimator and the Kasai autocorrelation estimator. For this purpose, the velocity estimator was averaged in time over sliding windows of 0.25s with a 75% overlap between two consecutive temporal windows (i.e. with an actual temporal step of 0.06s) and in space over all the voxels of the carotid artery (defined like in Figure 10), reflecting the global flow rate inside the vessel. Figure 11 shows that with the XDoppler and the Kasai autocorrelation estimators, the pulsatility of the artery can be followed during 3 to 4 cardiac cycles. The heart rate can then be estimated using the peaks of the XDoppler curve: in this example, the average time period is 1.35s, i.e. a heart rate of around 44bpm corresponding to the heart rate at rest of the volunteer when measured by an ECG, confirming the ability of the XDoppler estimator to reliably capture the flow dynamics, like the Kasai autocorrelation estimator is known to do.

Supplementary video shows a movie of the evolution of the axial blood velocity in the carotid artery and in the jugular vein during the same time period, allowing a dynamic volumetric visualization of the evolution of the flow dynamics inside the carotid artery and the jugular vein.

## 4. Discussion

In this work, we have introduced a novel phase-based velocity estimator derived from the XDoppler approach, which relies on the cross-correlation of orthogonal apertures of RCA arrays. We demonstrated that, thanks to its improved spatial resolution and contrast demonstrated in the seminal XDoppler paper, this estimator outperforms the conventional Kasai estimator with OPW in terms of accuracy of the flow estimation and robustness to aliasing.

By exploiting the temporal lag between transmissions with the rows and columns, a simple phase estimator of the axial velocity is obtained. This estimator is inherently biased and must be corrected using the complementary phase estimation between columns and rows. The combined XDoppler estimator is therefore based on the sum of the two cross-correlations of successive signals sampled by the orthogonal apertures of the RCA array, and can be interpreted as a translation of the "lag-1 autocorrelator" to RCA probe imaging – with the "lag 1" decomposed into two complementary half-lags with orthogonal spatial patterns.

The *in vitro* results confirm that the axial velocity can reliably be mapped in 3D at high frame rate using this new estimator and show that the enhancement of resolution and contrast of the XDoppler cross-correlation, which was initially introduced for power Doppler imaging, also improves volumetric color Doppler imaging. In the *in vitro* experiments, the Kasai estimator with OPW compounding systematically underestimated the velocity at the center of the tube and overestimated it around the edges of the pipe and also outside of it, where the velocity is supposed to be null. These features are directly linked to the fact that OPW technique has a large main lobe and strong sidelobes: the velocity in one voxel is averaged within the main lobe and with the lobes of the surrounding voxels, which acts like a low-pass filter for spatial frequencies, reducing the effective spatial resolution and therefore making it more difficult to correctly estimate both maximal and minimal velocities of the flow. Decreasing the main lobe dimension and the level of the sidelobes to compensate for this effect would imply focusing with more plane waves, reducing the effective framerate and the Nyquist velocity, which would induce aliasing. Thus, the optimal tradeoff between spatial and temporal resolution to obtain a correct velocity estimation is challenging to define with the autocorrelation estimator applied to OPW compounding. On the other hand, the XDoppler estimator allowed a better estimation of axial velocity profiles. It outperformed the traditional Kasai estimator thanks to its sharper spatial resolution, which provides a better sensitivity to slow flows – regarding very slow flows, the flow signal is filtered out by the spatiotemporal SVD filtering before the velocity estimation. This increased sensitivity to small velocities might be of interest for both tissue Doppler applications and smaller vessels blood flow characterization in functional ultrasound or rodent vascular imaging applications.

Additionally, both the *in* vitro and *in* vivo results showed that this new estimator outperformed the traditional Kasai estimator to estimate the velocity maxima thanks to a Nyquist velocity twice as high. This reduced sensitivity to aliasing effects may be particularly beneficial for estimating high magnitude velocities, such as in the case of carotid stenosis estimation or mitral valve regurgitation.

Although not investigated here, applying this estimator to other RCA compounding techniques that retain the phase information but offer sharper spatial resolution (e.g., RC-FMAS) could further enhance sensitivity to slow flows, while still being limited by the lower Nyquist velocity of the Kasai estimator for fast-flow measurements.

A potential limitation of the XDoppler velocity estimator is the integration time required to provide accurate estimates of the axial velocity. In this work, the convergence time was between 0.05 and 0.2s. This could be a limitation for imaging highly dynamic and pulsatile flows. Nevertheless, *in vivo* data on human carotid artery confirm that XDoppler velocity estimator can better image the flow velocity and its dynamics, particularly due to its reduced sensitivity to aliasing effects. These results also suggest that this method could evaluate blood flow dynamics in various phases of the cardiac cycle despite its intrinsic integration time. These features could be of particular interest for volumetric vascular imaging in complex flow environments such as carotid stenosis monitoring or cardiac imaging.

Another technical limitation is the potential probe heating: as described in the subsection dealing with calibration of the ultrasound sequence, we used a sequence that comprised pauses, enabling it to meet the IEC heating standards. To design completely continuous acquisitions meeting these requirements, it would be necessary to decrease the acoustic power.

Another limitation of this technique is the potential for signal decorrelation during the compounding operation in the case of rapidly moving scatterers [46]. Although we did not observe this issue in our experiments, if it arises, various techniques of motion compensation could be investigated to tackle this problem [57], [58].

Although the Nyquist limit of the XDoppler estimator is structurally higher than the Kasai estimator Nyquist velocity, it still relies on a tradeoff between the pulse repetition frequency, the number of plane waves used for synthetic focusing, and the central frequency of the pulse. In certain rapid flow imaging contexts (cardiac imaging, aortic Doppler, carotid stenosis, …), it might not be sufficient to capture the highest velocities. In such cases, specific de-aliasing techniques (such as interleaving or repeating emissions as in [35], [59], [60], [61], [62], creating a « staggered » scheme as in [57], or taking advantage of the broadband nature of the signals as in [63]) could be investigated to extend the Nyquist limit while preserving the sensitivity of the XDoppler velocity estimator to slow flows.

Since the beam-to-flow angle is visibly close to 90° in our *in vivo* experiments, we did not perform an angle correction to estimate the total magnitude of the velocity vectors inside the artery and the vein, because the uncertainty level would be too high: estimating the total magnitude of the velocity vectors could be done by performing vector flow imaging, which would enable to resolve any beam-to-flow angle in the elevational and azimuthal directions. Given the recent developments in vector flow imaging with row-column array probes, based either on the transverse oscillation tensor estimator [22], [23], [24], [64] or the autocorrelation estimator [26], further work could indeed focus on integrating the XDoppler estimator into a vector flow plane wave imaging scheme to leverage on its ability to resolve both fast and slow flows. Such applications could be valuable for evaluating volumetric wall shear stress in arteries [65], this quantity being of paramount importance in many biofluid mechanics problems [66].

## 5. Conclusion

In this study, we developed and validated a novel phase-based velocity estimator for row-column addressed (RCA) ultrasound probes by extending the concept of XDoppler imaging scheme, initially developed for power Doppler imaging, to the estimation of axial phase velocity for color Doppler imaging. Thanks to its better spatial and temporal resolution compared to conventional estimators, this new estimator could be useful to overcome the challenges associated with 3D imaging and velocity estimation in complex flow and tissue motion environments.

**Acknowledgements**

We thank Yinshuang Zhou for her internship work. We acknowledge the French Ministère de l'Enseignement Supérieur et de la Recherche for their funding through an Allocation Moniteur Polytechnicien grant, as well as INSERM (Institut national de la santé et de la recherche médicale) for their financial support.

**Data statement**

Data supporting the findings associated with this study are available upon request to the authors.

## Figures

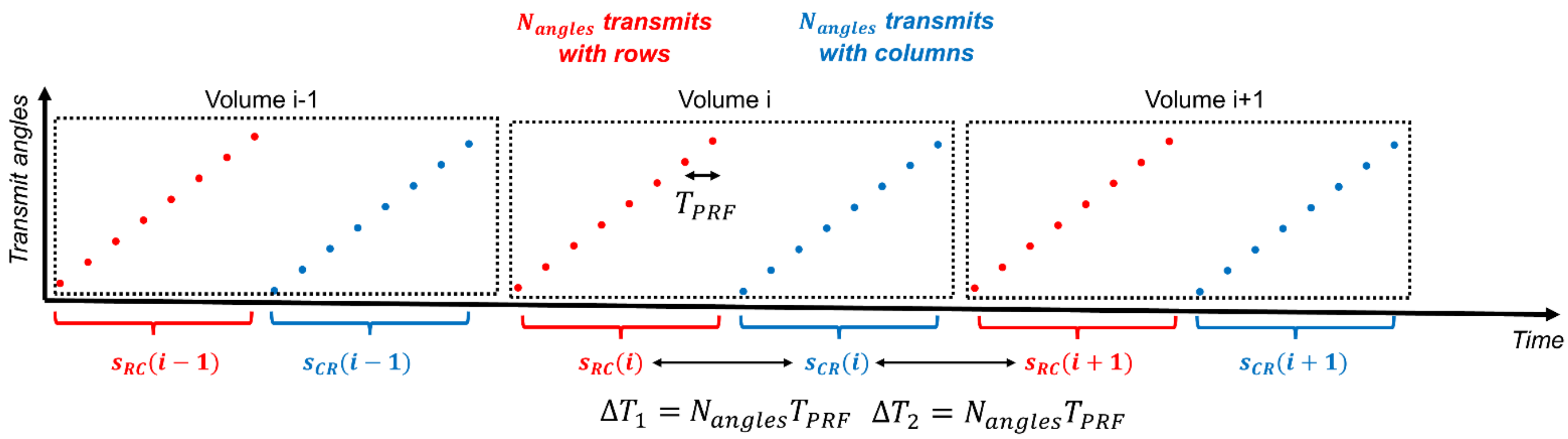

Figure 1. Row-column plane wave imaging sequence

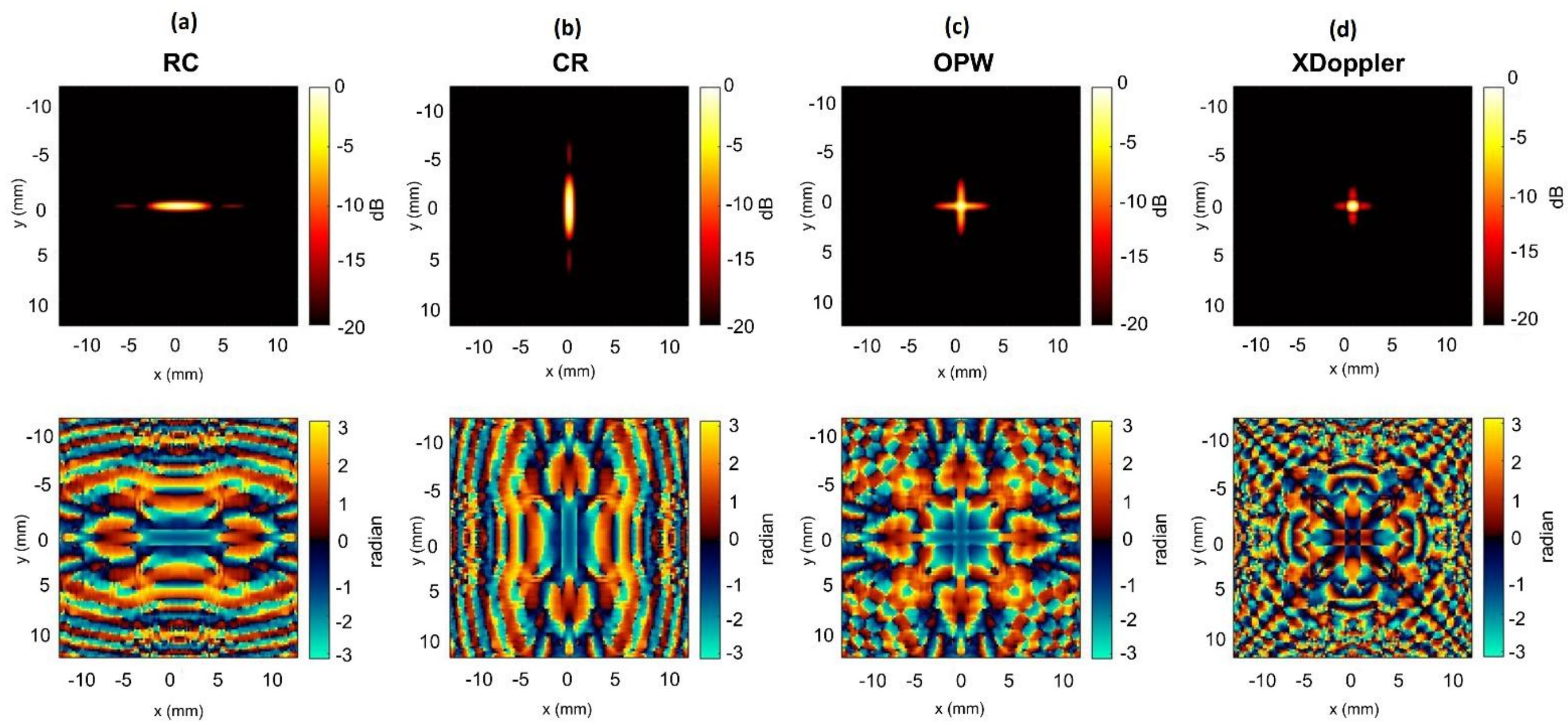

Figure 2. Simulation of the complex signals (top: amplitude/bottom: phase) of the Point Spread Functions of RC, CR, OPW and XDoppler compounding for a single scatterer in the focal plane (F#1)

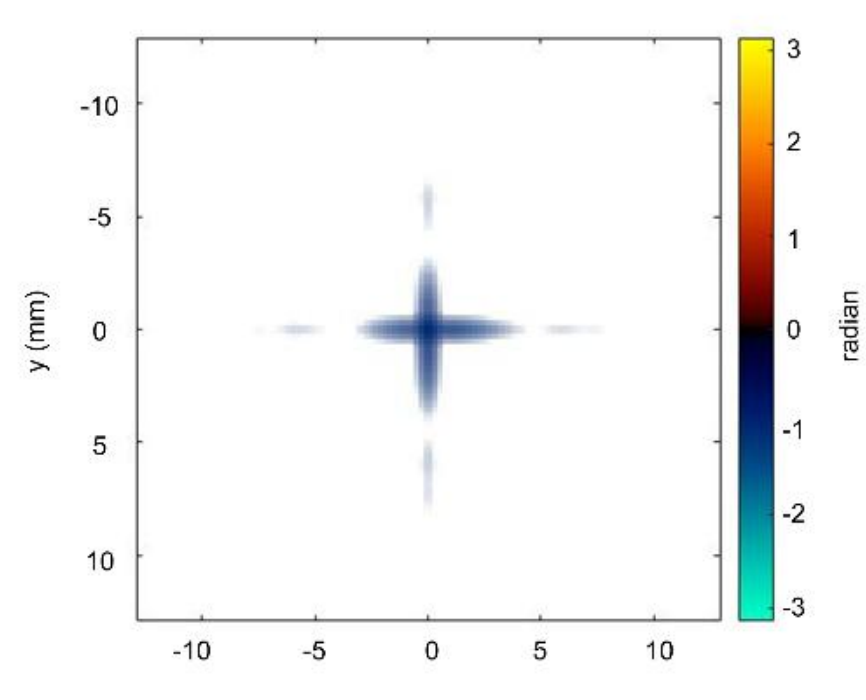

Figure 3. Phase of the autocorrelation of the complex signals echoed by a scatterer initially positioned in (0,0) and moving away from the probe with a constant axial velocity $v_z$=12mm/s using OPW compounding. The ultrasound sequence is the same as in Figure 1. The Nyquist velocity is $v_N$=46.2mm/s. The correlation is averaged over a duration

of 500ms before phase computation. The phase is weighted by the amplitude of the OPW PSF, with a dynamic range of 25dB to improve the visualization of the phase inside and around the main lobe.

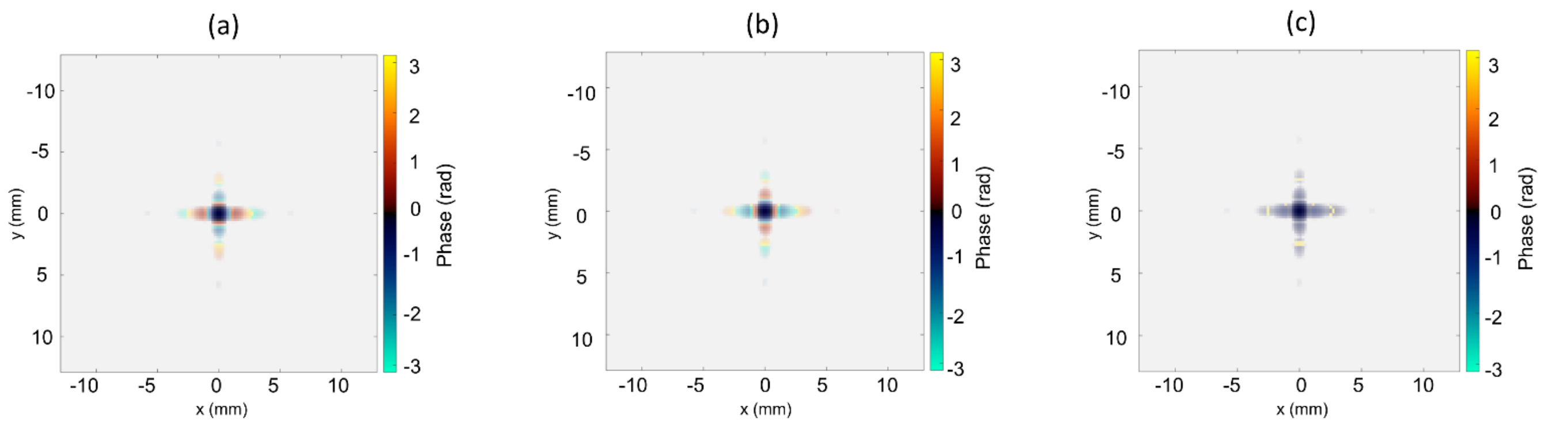

Figure 4. Phase patterns of cross-correlations based on the same simulation sequence and data as in Figure 1 and Figure 3. (a) Phase of the cross-correlation of two consecutive PSFs $s_{RC}(i)s^*_{CR}(i+0.5)$ for a scatterer moving away from the probe with a velocity $v_z$ = 12mm/s (b) Phase of the cross-correlation of two consecutive PSFs : $s_{CR}(i+0.5)s^*_{RC}(i+1)$ for a scatterer moving away from the probe with a velocity $v_z$ = 12mm/s (c) Sum of the phases of $s_{RC}(i)s^*_{CR}(i+0.5)$ and $s_{CR}(i+0.5)s^*_{RC}(i+1)$ for a scatterer moving away from the probe with a velocity $v_z$ = 12mm/s. All the correlations are averaged over a duration of 500ms before phase computation. The phase is weighted by the amplitude of the XDoppler PSF, with a dynamic range of 25dB to improve the visualization of the phase inside and around the main lobe.

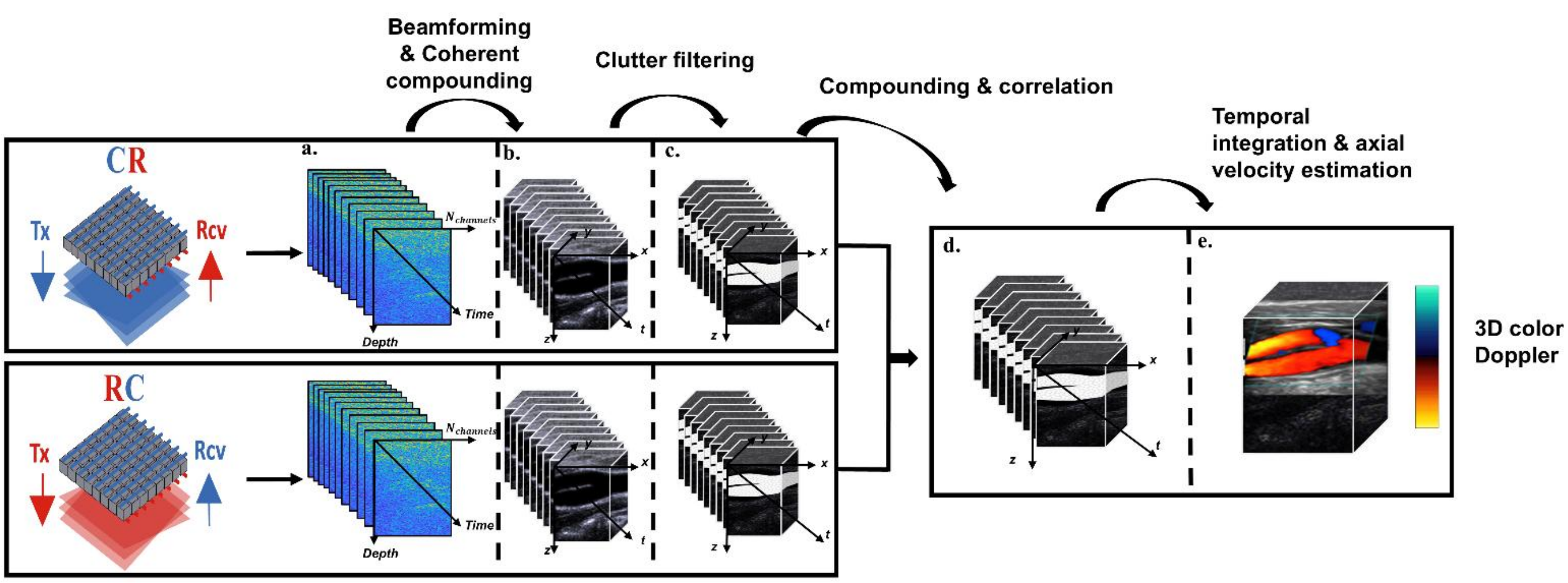

Figure 5. Processing pipeline for XDoppler velocity estimation

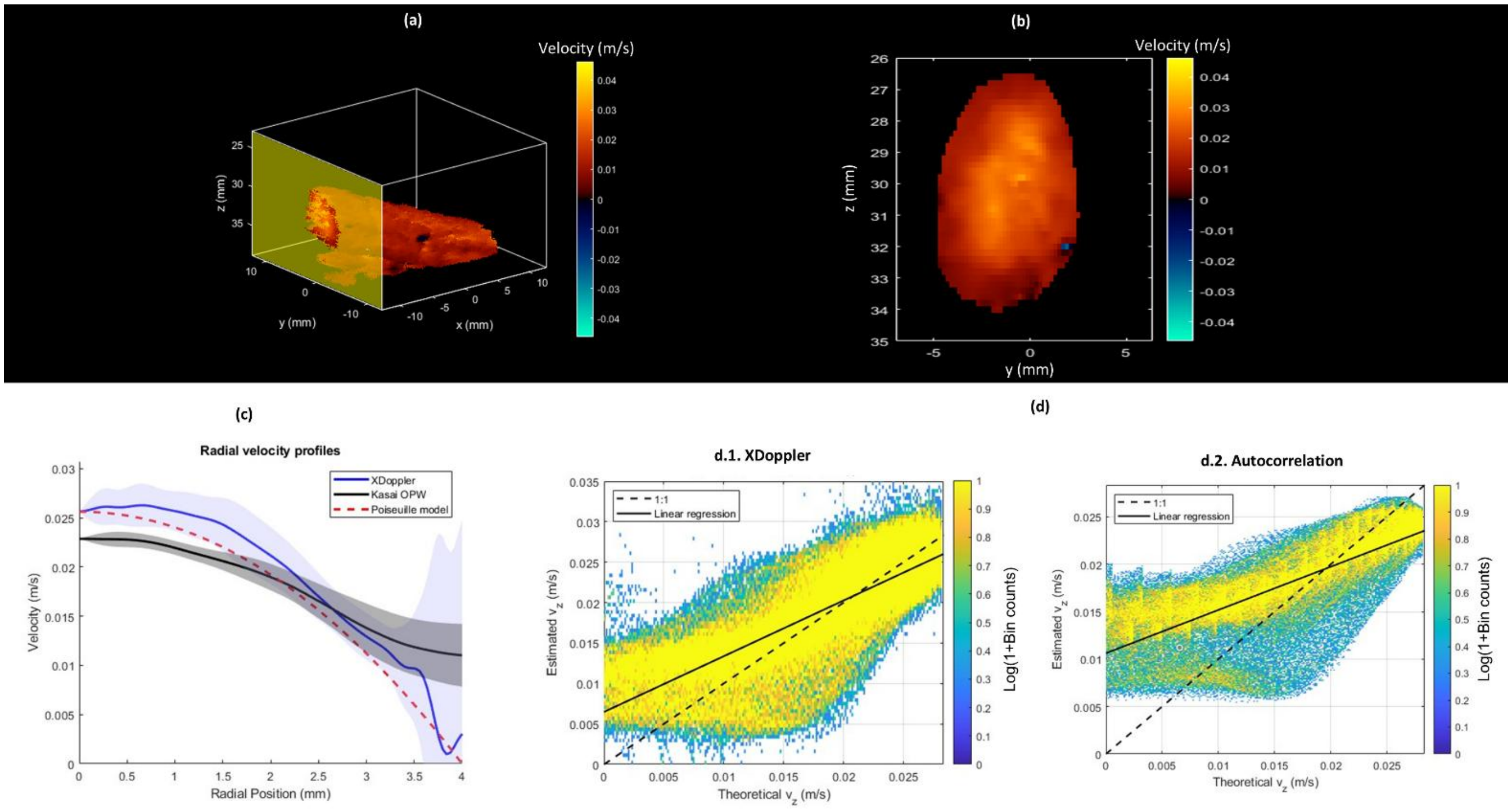

Figure 6. (a) Volumetric representation of XDoppler velocity estimator for a laminar flow in a pipe (b) 2D slice of XDoppler velocity estimator for a laminar flow in a pipe (c) Comparison of radial velocity profiles of XDoppler and Kasai estimations with a Poiseuille flow model. The continuous line represents the mean velocity value and the error bars represent the standard deviation over all the selected radii (d) Estimated velocity estimates distributions compared to a Poiseuille velocity profile. The colorbar represents the bin counts in logarithmic scale : (d.1) XDopppler estimator (d.2) Kasai autocorrelation estimator

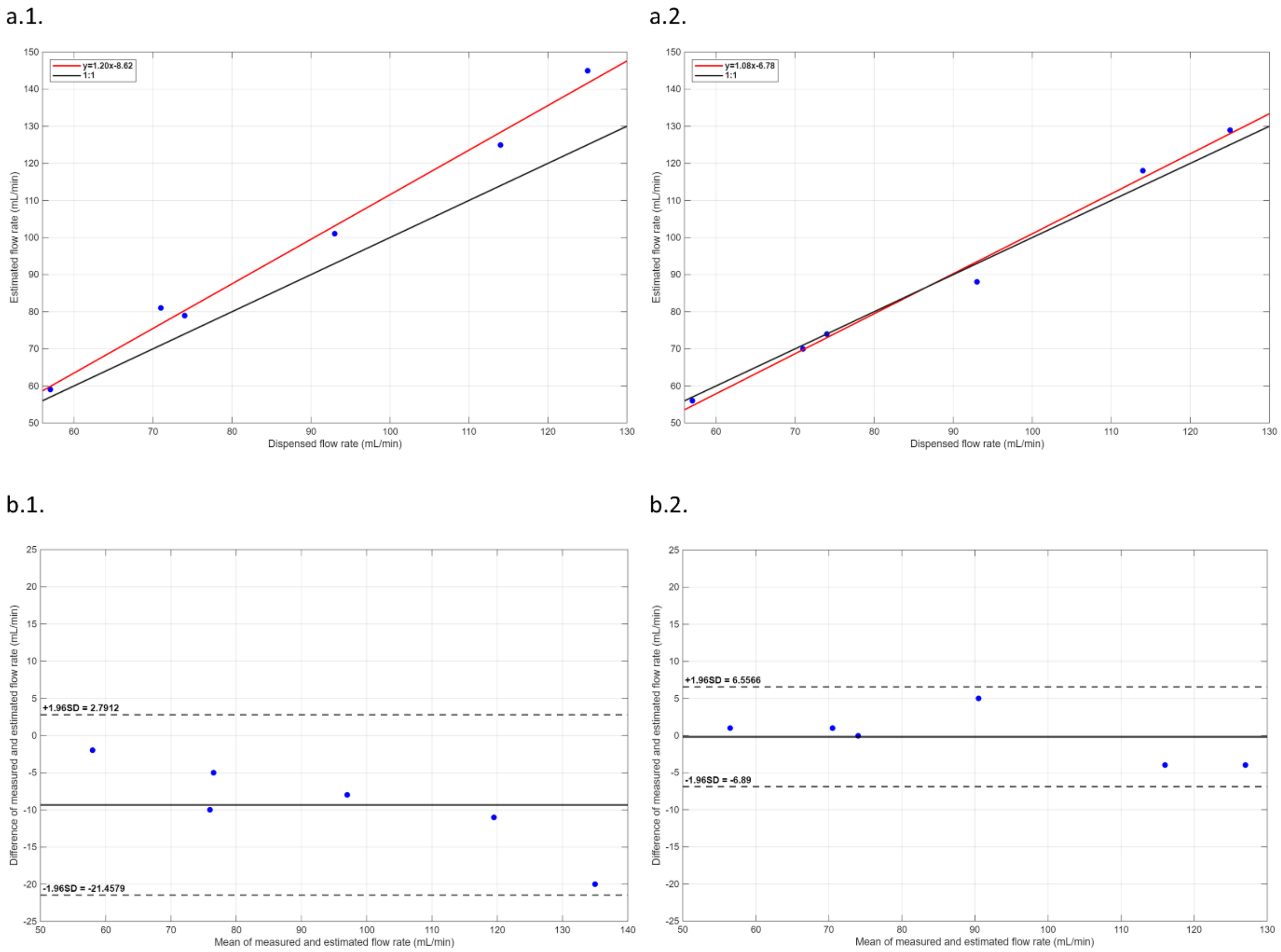

Figure 7. Comparison of the quality of the agreement between estimated flow rates and measured flow rates between Kasai estimator and XDoppler estimator for N=6 flow rates in experiment 1: a.1. Correlation – OPW autocorrelation method. a.2. Correlation– XDoppler method. b.1. Bland-Altman plot– OPW autocorrelation method. b.2. Bland-Altman plot– XDoppler method.

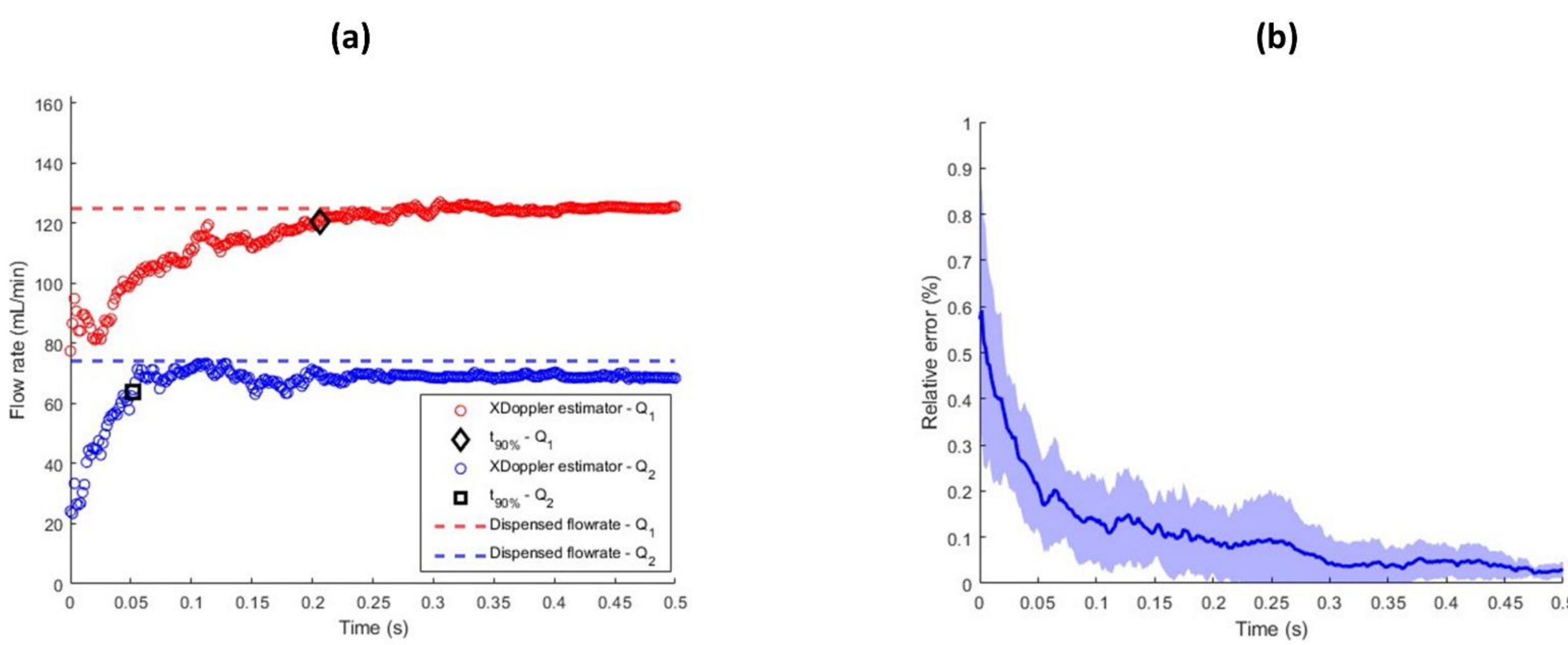

Figure 8. Temporal evolution and convergence time of the XDoppler estimator: (a) Evolution of the flow rates estimations with time for two different flow rates ($Q_1 = 125mL/min$ and $Q_2 = 74mL/min$) of experiment 1. The evolution is similar for the 4 other flow rates, which are not represented for readability purposes (b) Evolution of the mean relative error and its standard deviation for N=6 different flow rates of experiment 1

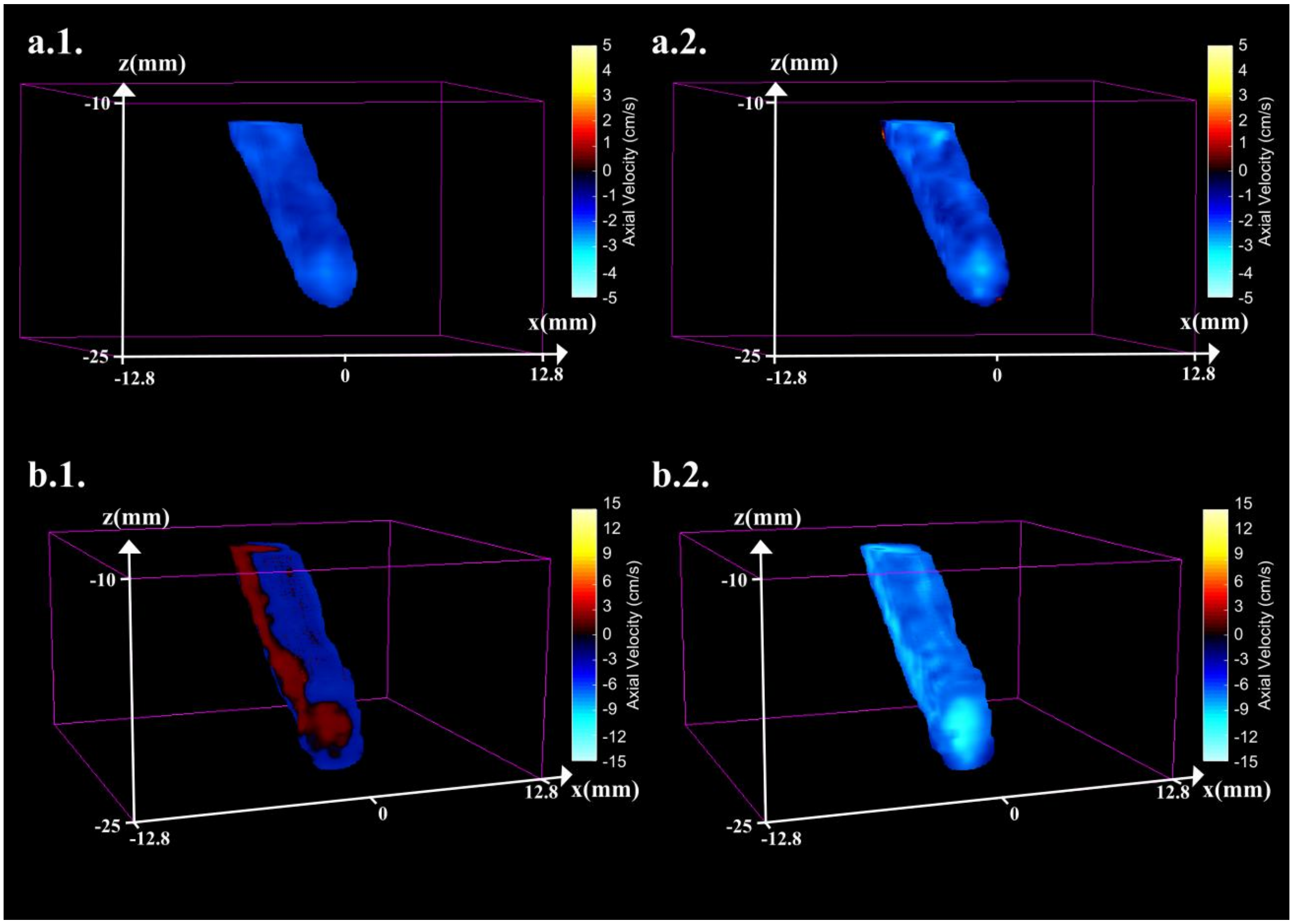

Figure 9. Evaluation of the robustness to aliasing in experiment 2. a: Volumetric axial velocity maps for a low flow rate ($v_z = 3cm/s$) computed with autocorrelation velocity estimator based on OPW compounding (a.1.) and XDoppler velocity estimator (a.2). No aliasing is observed, and the velocity values are rather similar with the two estimators. b: Volumetric velocity maps for a high flow rate ($v_z = 12cm/s$) computed with autocorrelation estimator (b.1.) and XDoppler estimator (b.2). There is no aliasing in XDoppler estimation strong aliasing occurs with the autocorrelation estimator. All the velocity maps are masked by applying a threshold on the power Doppler to keep voxels with a XDoppler dynamic superior to 15 dB for visualization purposes

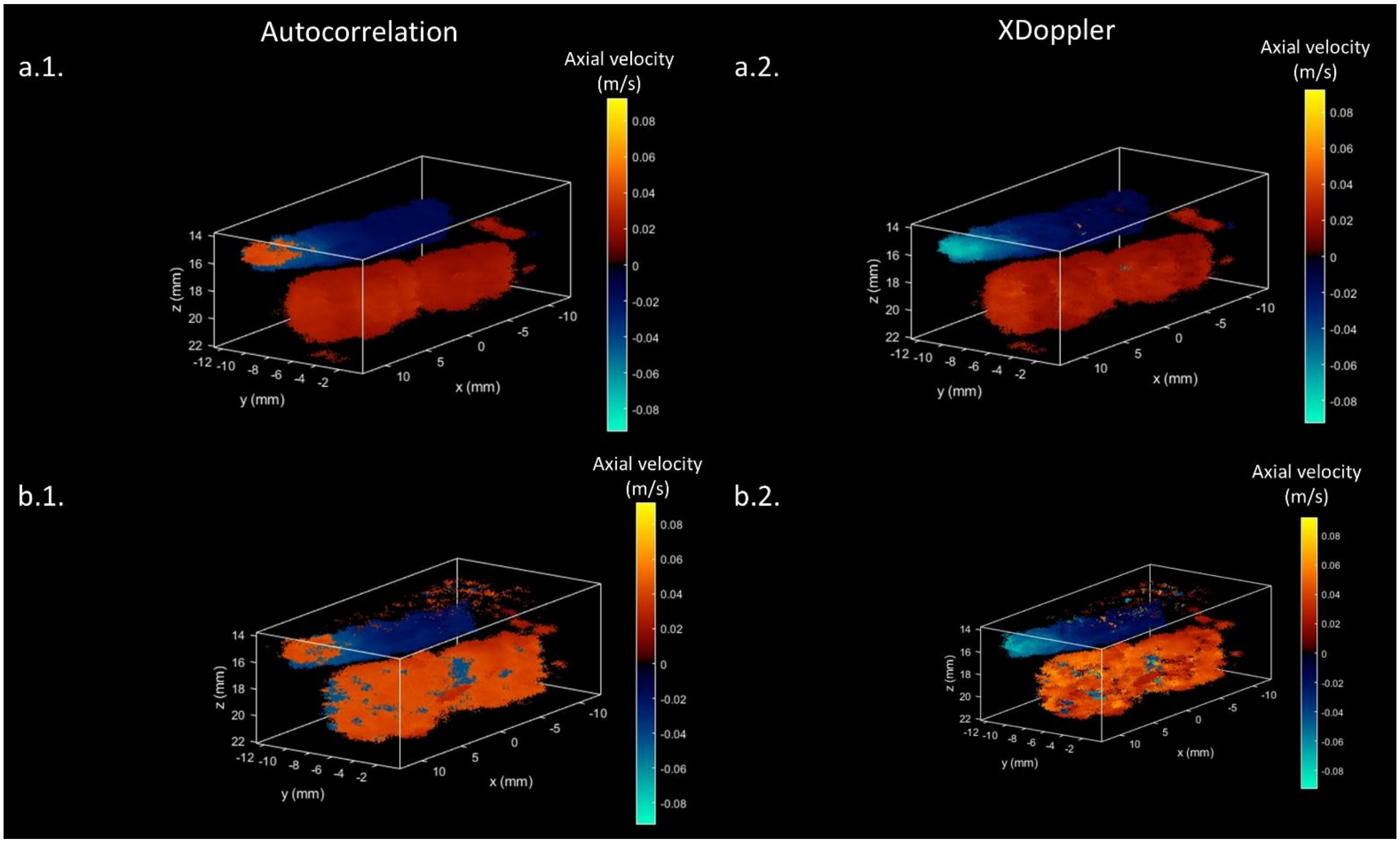

Figure 10. *In vivo* estimation of axial blood flow velocity in a healthy volunteer human carotid at several steps of the cardiac cycle (a.1) OPW compounding with autocorrelation estimator during diastole (a.2) XDoppler estimator during diastole (b.1) OPW compounding with autocorrelation estimator during systole (b.2) XDoppler estimator during systole

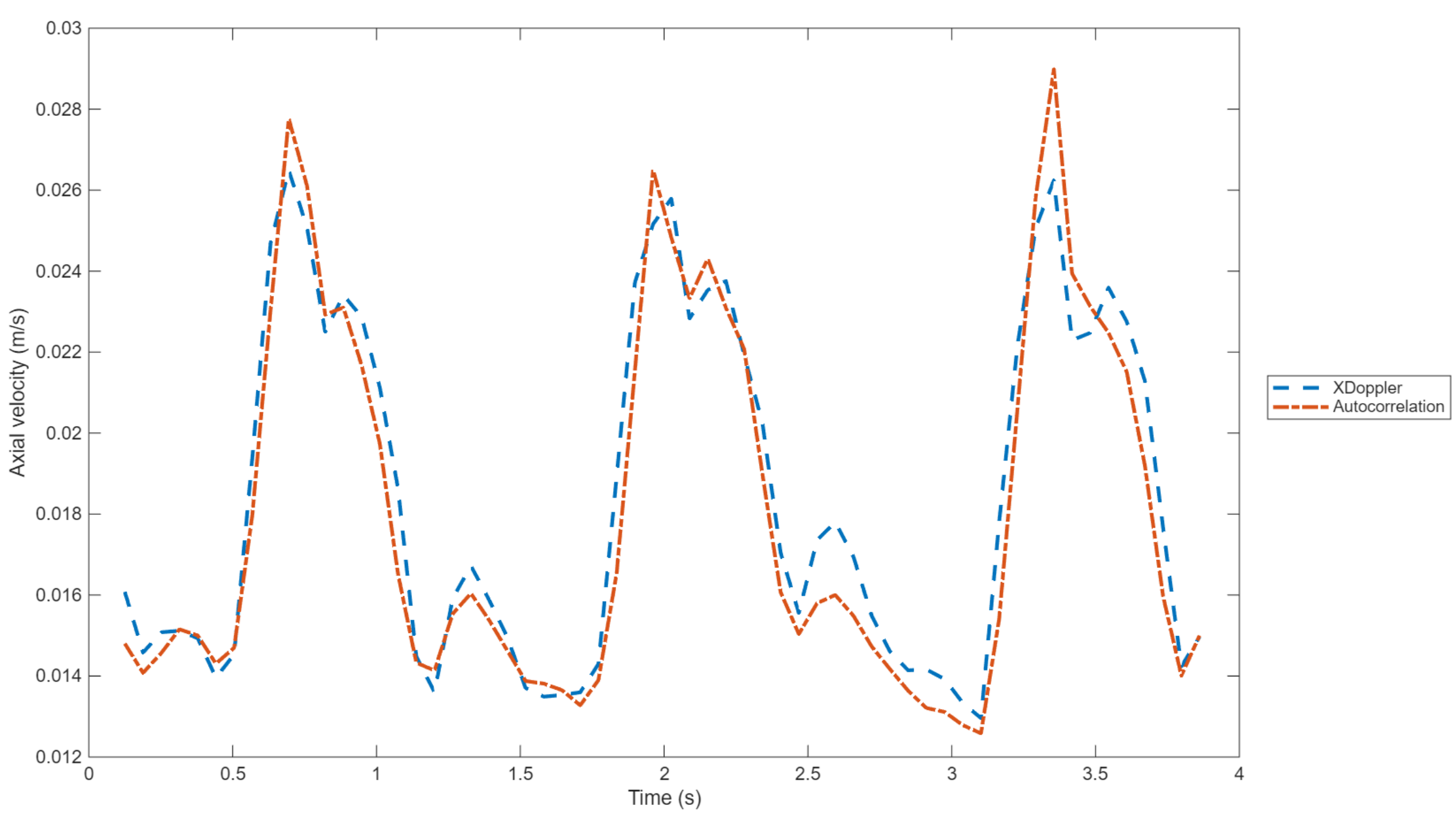

Figure 11. Estimation of the mean axial velocity inside the carotid over time using the XDoppler and the Kasai autocorrelation estimators